\documentclass[12pt]{article}

\newcommand{\be}{\begin{equation}}
\newcommand{\ee}{\end{equation}}
\newcommand{\bea}{\begin{eqnarray}}
\newcommand{\eea}{\end{eqnarray}}

\newcommand{\bit}{\begin{itemize}}
\newcommand{\eit}{\end{itemize}}
\newcommand{\no}{\noindent}

\begin{document}

{\sf 

\title{Ten and eleven dimensional perspectives on N=2 black holes}
\author{Ansar Fayyazuddin\footnote{email: Ansar\_ Fayyazuddin@baruch.cuny.edu}}

\maketitle

\begin{center}

{\it Department of Natural Sciences, Baruch College, City University of New York, New York, NY 10010}

\end{center}

\vspace{1cm}

\begin{abstract}
We give an 11 and 10 dimensional supergravity description of M5-branes wrapping 4-cylces in a Calabi-Yau manifold and carrying momentum along a transverse S$^1$.  These wrapped branes descend to a class of N=2 black holes in 4 dimensions.  Our description gives the conditions on the geometry interpolating between the asymptotic and near-horizon regions.  We employ the ideas of geometric transitions to show that the near horizon geometry in ten dimensions is AdS$_2\times$S$^2\times CY_3$ while in 11 dimensions it is AdS$_3\times$S$^2\times CY_3$.  We also show how to obtain the complete N=2 black hole supergravity solution in 4 dimensions for this class of black holes starting with our 11-dimensional description.  Finally, we generalize our results on the 10 and 11 dimensional near horizon supergravity solution to the case of black holes carrying arbitrary charges (D0-D2-D4-D6 in the type IIA description).  We argue that the near horizon geometry corresponding to wrapped D6 and D2 branes in 11 dimensions is AdS$_2\times$S$^3\times CY_3$.
 \end{abstract}

\vspace{-20cm}
\begin{flushright}
BCCUNY-HEP /06-01 \\
hep-th/0603141
\end{flushright}

\thispagestyle{empty}

\newpage

\tableofcontents

\section{Introduction}
The subject of N=2 black holes is usually formulated in the context of d=4, N=2 supergravity \cite{n2}.  This theory arises as the low energy description of type II string theory on compact Calabi-Yau 3-folds where the energies are assumed to be too low to resolve the geometry of the Calabi-Yau.  It is our purpose to inject some new perspectives in this field by going beyond this approximation.  This allows us to address questions concerning the geometry transverse to the uncompactified 4-d space-time.  

In principle there is no restriction on the size of Calabi-Yau manifolds in string theory.  If the Calabi-Yau is non-compact or if we are probing at energies sufficiently large to resolve the geometry of the Calabi-Yau, a d=4 supergravity description may be a bad approximation.  Actually, the last decade has placed non-compact Calabi-Yau manifolds at center stage in string theory.  The singularity structure of local Calabi-Yau manifolds have provided us with the setting for much interesting physics, while global properties have occupied a less prominent position.  It is for this reason that we attempt to describe N=2 black hole geometry in eleven and ten dimensions and recover the d=4 description as a low-energy approximation of the black holes we describe.  

Our construction is based on modifying what is known about the geometry of M5-branes wrapping 4-cycles in Calabi-Yau 3-folds \cite{amherst}.  Our modification adds momentum along the M5-branes transverse to the Calabi-Yau.  These solutions descend to N=2 black holes in four dimensions.  After determining general properties of these solutions in d=11 and the reduction down to d=10, we proceed to determine the near horizon geometry in 10 and 11 dimensions.  We also find explicit d=4 solutions for the black holes and generalize our d=10 and d=11 near horizon results to the more general case where D2 and D6 branes are also present.  The study of these black holes was spearheaded in \cite{behrndt} where some of the results below were anticipated.  The microscopic counting of these black holes was carried out in \cite{vafacy} and \cite{msw}.  The macroscopic entropy was studied in detail in \cite{behrndt} and \cite{ms}, and extended to include higher-derivative corrections in \cite{cardoso}.  

Recently a number of new issues have re-animated the subject of N=2 black holes.  The most important of these addresses the question of calculating the entropy of black holes in theories with higher derivative terms beyond the Einstein action.  These are leading terms for black holes with vanishing horizon area and provide possible resolutions to various puzzles.  In this paper we will not have anything to say on this subject but we hope that the ideas presented here will have something to teach us about these questions as well.  

Note added: As I was finishing this work, I became aware of \cite{ssty} which also considers the near horizon geometry of N=2 black holes in 10 dimensions and overlaps with sections 3 and 4.  

\section{Wrapped M5-branes and M-waves}
In this section we construct the eleven dimensional geometry of M5-branes wrapping 4-cycles with momentum along one direction.  These black holes were originally studied macroscopically and microscopically in \cite{msw}.

Our starting point is the metric for M5-branes wrapping holomorphic 4-cycles in Calabi-Yau manifolds.  The eleven dimensional supergravity solution was first constructed in \cite{amherst}.  The metric is:
\be
ds^2 = H^{-1/3}(-dt^2 + dy^2) + H^{2/3}\sum_{a=1}^{3}(dx^a)^2 + 2H^{-1/3}g_{m\bar{n}}dz^mdz^{\bar n}. \label{m5}
\ee
where $g_{m\bar{n}}$ is a K{\" a}hler metric \cite{amherst,Tasneem} on a 3-fold with coordinates $z^m$, $g$ depends on the transverse coordinates $x^a$ as well.  The determinant of the metric $g$ is related to $H$ as follows \cite{amherst}: 
\be
det(g) = aH^2|h|^2, \label{H}
\ee
where $h$ is holomorphic in $z^m$ and $a$ is a constant.  In other words $H^2$ is the determinant of $g$ up to a Jacobian for a holomorphic change of coordinates.  The 4-form field strength of 11-dimensional supergravity is expressed in terms of the metric and $H$ as follows:
\bea
F_{123m} &=& \frac{i}{2}\partial_mH \nonumber \\
F_{abm\bar n}&=&\frac{i}{2}\epsilon_{abc}\partial_cg_{m\bar{n}} \label{F}
\eea
Here $\epsilon_{123}=1$ is a completely antisymmetric symbol in the overall transverse directions.
The condition that $g$ is K{\" a}hler guarantees that $d*F=0$ which is the Bianchi identity when only M5-branes are present since they couple magnetically to the 3-form potential.  In addition we must satisfy the source equation $dF = "0"$ where $"0"$ represents the delta function source with support on the 4-cycle in the Calabi-Yau.  In what follows we will replace $"0"$ with $0$ but we should keep in mind that there is a source term present.  This equation of motion can be cast in the form:  
\be
\nabla_\perp^2g_{m\bar{n}} + 2\partial_m\partial_{\bar n}H =0.\label{eom}
\ee
Here $\nabla^2_\perp$ is the flat Laplacian in the transverse directions $1,2,3$.  Since $Ha|h|^2 = det(g)$, this is a non-linear partial differential equation.  

This supergravity solution is supersymmetric and is invariant under supersymmetry transformations involving variation parameters satisfying\cite{Tasneem}:
\bea
\epsilon &=& \alpha + \beta \nonumber \\
\Gamma_m\alpha &=& \Gamma_{\bar m}\beta = 0\nonumber \\
\Gamma_{0y}\alpha &=& H^{-1/3}\alpha \\
\Gamma_{0y}\beta &=& -H^{-1/3}\beta \nonumber \label{susy}
\eea

We will now introduce momentum in the $y$ direction in the above solution.  We would like to do this without breaking any further supersymmetry.  In other words we would like the supersymmetry variation parameters under which our solution will be invariant to continue to be the same as (\ref{susy}).  We are lead to the ansatz:
\be
ds^2 = H^{-1/3}(-g_{00}dt^2 + g_{yy}(dy + Adt)^2) + H^{2/3}\sum_{i=1}^{3}(dx^i)^2 + 2H^{-1/3}g_{m\bar{n}}dz^mdz^{\bar n}.
\ee
Where all metric components are independent of $t,y$.  
The requirement that this metric satisfy the Killing spinor equation for 11-d supergravity:
\be
\delta_{\epsilon}\Psi_{I} = (D_I
+ \frac{1}{288}{\Gamma_{I}}^{ABCD}F_{ABCD}
-\frac{1}{36}\Gamma^{ABC}F_{IABC})\epsilon = 0
\ee
with $F$ given by (\ref{F}) gives us the constraints:
\bea
d\log(g_{00}g_{yy}) &=& 0 \nonumber \\
d A &=& d g_{00}.
\eea
Thus, supersymmetry requires that the metric take the form:
\be
ds^2 = H^{-1/3}(-f^{-1}dt^2 + f(dy + (f^{-1}-1)dt)^2) + H^{2/3}\sum_{i=1}^{3}(dx^i)^2 + 2H^{-1/3}g_{m\bar{n}}dz^mdz^{\bar n}. \label{bh11}
\ee

The matter fields (a.k.a. $F$) are the same in the solutions before and after the momentum is added.  We have to make sure that the metric (\ref{bh11}) satisfies the equations of motion.  This requirement gives us a constraint on $f$:
\be
H^{-1}\nabla^2_{\perp}f + 2g^{m\bar{n}}\partial_m\partial_{\bar n} f = 0. \label{f}
\ee

To summarize, the metric (\ref{bh11}) and the equations (\ref{H}), (\ref{f}) and (\ref{eom}) along with (\ref{F}) give us the black hole solution in eleven dimensions that we will study in the remainder of this paper. 

We now give a ten dimensional description of the above black hole in type IIA supergravity.   Since we have translational symmetry in the $y$ direction we take it to have $S^1$ topology and reduce to the spacetime transverse to it.  We extract the $10$ dimensional string frame metric $ds^2_{10}$ using:
\be
ds^2 = e^{4\phi /3}(dy + Adt)^2 + e^{-2\phi /3}ds^2_{10}.
\ee 
The complete solution in ten dimensional type IIA supergravity is then:
\bea
ds^2_{10} &=& -H^{-1/2}f^{-1/2}dt^2 + H^{1/2}f^{1/2}(dr^2 + r^2d\Omega^2) + 2H^{-1/2}f^{1/2}g_{m\bar n}dz^m dz^{\bar n}  \nonumber \\
C_1 &=& A dt = (f^{-1}-1)dt \nonumber \\
e^{\phi} &=& H^{-1/4}f^{3/4} \\
dC_3 &=& F \nonumber \label{iia}
\eea 
with $F$, $f$ and $H$ given by (\ref{F}), (\ref{f}), (\ref{eom}) and (\ref{H}) satisfying the constraints as above.

\section{The ten dimensional near-horizon solution of N=2 black holes}
Black holes interpolate between two distinct regimes - the near-horizon region and the asymptotic space-time far from the black hole.  In the asymptotically far region we approach the vacuum - $R^{3,1}\times CY$.  In this section we would like to explore the near-horizon region and discover some facts about its geometry.

Maldacena has defined \cite{maldacena} the near-horizon region as a low-energy truncation of space-time.  This definition is a very powerful way of systematically taking a given black hole solution and separating out the near-horizon region.  There is an alternative view which incorporates the lessons of all known near-horizon limits obtained using Maldacena's method.  This view was first advocated by Vafa \cite{vafa} who argued that the near-horizon region is characterized by a large N geometric transition where physical localized branes are replaced by fluxes through finite sized cycles.  We will follow this latter characterization of the near-horizon region to determine a space-time that satisfies all of our equations but does not involve localized branes, only fluxes.

Recall that we are studying a system of D4-branes wrapping 4-cycles inside of the Calabi-Yau and D0-branes on a 0-cycle.  If we follow Vafa's characterization of the near-horizon limit then we would expect that  in this limit, if we denote by $G$ the field strength associated with a gauge potential which couples electrically to one of the branes, then $*G$ should be a form whose integral over an appropriate cycle calculates the flux produced by the branes but now in the absence of any localized sources.

Following Vafa we propose that the near-horizon limits of the field strengths $F = *G_6 = *dC_5$ and $G_8 = *dC_1= *df^{-1}$ should calculate appropriate fluxes through finite sized cycles.  From now on to distinguish near-horizon quantities from the full supergravity solution we will add a prime ' to all quantities evaluated in the near-horizon limit.  Since the D4-brane wraps a 4-cycle in $M$ we will take the flux produced by the D4-brane to be through a 4-cycle that is a product of the dual 2-cycle in $M$ and a 2-sphere in the transverse space.  For the D0-brane we will take the 8-cycle to be a product of $M$ and a transverse 2-sphere.  We expect all of these cycles to be of finite size.

Before taking the near-horizon limit, let us recall that:
\bea
F &=& -\frac{i}{2}r^2\partial_r g_{m\bar{n}}dz^m\wedge dz^{\bar n}\wedge dvol(S^2) 
+ \frac{i}{2}r^2\partial_mH dz^m\wedge dr\wedge dvol(S^2) \nonumber \\
&-&  \frac{i}{2}r^2\partial_{\bar m}H dz^{\bar m}\wedge dr\wedge dvol(S^2) \nonumber \\
G_2 &=& dC_1 = \partial_r f^{-1} dr\wedge dt + \partial_m f^{-1}dz^m\wedge dt + \partial_{\bar m} f^{-1}dz^{\bar m}\wedge dt \label{generalF}
\eea
Where $dvol(S^2)$ denotes the volume form of a unit 2-sphere.  In the near horizon limit:
\bea
F&\rightarrow& F'=\omega_2\wedge dvol(S^2) \nonumber \\
G_8 &\rightarrow& G_8'= \omega_6\wedge dvol(S^2) \label{nearhF}
\eea
where $\omega_2$ and $\omega_6$ are a 2- and 6-form, respectively, on the 6-dimensional manifold $M'$.  We will assume that neither $\omega_2$ nor $\omega_6$ depends on $r$ in the near-horizon limit.   This is because their integrals compute the flux produced by the original localized D-branes at $r=0$ and the fluxes are thus independent of $r$.  

Comparing the general expressions (\ref{generalF}) to the near horizon ones (\ref{nearhF}) we discover that we must take $\partial_mH'=0$ and $\partial_m f' =0$.  With these restrictions on $H'$ and $f'$ our master equations (\ref{eom}) and (\ref{f}) take the form:
\bea
\frac{1}{r^2}\partial_r(r^2\partial_r g_{m\bar n}' )&=& 0 \nonumber \\
\frac{1}{r^2}\partial_r(r^2\partial_r f') &=& 0 \label{nearhlaplacian}
\eea
Let us define the near horizon K{\" a}hler form $J' = ig'_{m\bar{n}}dz^m\wedge dz^{\bar n}$ then (\ref{nearhlaplacian}) and (\ref{nearhF}) imply   
\be
J' = \frac{1}{r}\omega_2.
\ee
Since $g$ is a K{\" a}hler metric so is the near horizon $g'$, implying that $d_6J' =0$.  With this restriction we have that:
\be
d_6\omega_2 =0
\ee

We are now in a position to calculate $H'$.  Recall that $H^2a^2|h|^2 = \det {g}$ where $h$ is holomorphic in $z^m$ and $a$ is a constant.  This identity is equivalent to the relation:
\be
Ha|h|dz^1\wedge dz^{\bar 1}\wedge dz^2\wedge dz^{\bar 2}\wedge dz^3\wedge dz^{\bar 3} = \frac{1}{6}J\wedge J\wedge J
\ee
where $J$ is the K{\" a}hler form associated with $g_{m\bar n}$ on $M$.  First we fix $ah$.  Consider the limit $r\rightarrow \infty$.  In this limit we should get the asymptotic space time $R^{3,1}\times CY$, or $H\rightarrow 1$ and  $J \rightarrow J_{CY}$.  Thus we find that  
\be
a|h|dz^1\wedge dz^{\bar 1}\wedge dz^2\wedge dz^{\bar 2}\wedge dz^3\wedge dz^{\bar 3} = dvol(CY).
\ee
Thus we have the identity:
\be
Hdvol(CY) = \frac{1}{6}J\wedge J\wedge J \label{id}
\ee
This identity is general and valid through out the black hole space time.  Let us consider the near horizon limit of (\ref{id}):
\be
H'dvol(CY') = \frac{1}{6}J'\wedge J'\wedge J' \label{id'}
\ee
We are used to thinking of the near-horizon geometry as the $r\rightarrow 0$ limit from the 4-dimensional point of view.  In the present case it also acts on the remaining six dimensional part as well.  In fact, thinking of the near-horizon limit as was originally defined in \cite{maldacena} as a low-energy decoupling limit we expect it to act on the Calabi-Yau at $r\rightarrow \infty$ as well. 
Here by $CY'$ we mean the vacuum Calabi-Yau in this limit .  As noted above, $H'$ does not depend on $z^m$, the coordinates on $M$. If $CY'$ is compact then we can integrate both sides to get 
\be
H' V'_{CY} = \frac{1}{6r^3}\int \omega_2^3.
\ee
where $V'_{CY}$ is the volume of the Calabi-Yau manifold $CY'$.  So finally we have,
\be
H' = \frac{1}{6V'_{CY}}\frac{1}{r^3}\int\omega_2^3. \label{H'}
\ee
It is not necessary for the $CY'$ to be compact, however.  The relation (\ref{id'}) ensures that in general 
\be
H' = \frac{c}{r^3}
\ee
where $c$ is a constant.  

Notice that equation (\ref{id'}) implies that $J'$ (or $\omega_2$) is a K{\"a}hler form on a {\it Calabi-Yau} manifold.  This is because the Ricci tensor associated with the K{\" a}hler form $J'$ is proportional to $\partial_m\partial_{\bar n}\ln H'$ which vanishes identically due to the constraint $\partial_m H'=0$, valid in the near-horizon limit.  

Next we find $f'$ by solving (\ref{nearhlaplacian}):
\be
f' = \frac{a_0}{r}.
\ee
We have now found expressions for all the quantities we wanted.

We are finally done with determining all quantities in the supergravity solution.  We summarize our results below and examine the solution in 10 and 11 dimensions.

\section{Near horizon supergravity solutions in 10 and 11 dimensions}
In the previous section we determined all the quantities we needed to write down the complete supergravity solution in the near-horizon region in 10 and 11 dimensions.  We assemble these results here away from the clutter of the last section.

Using the results from the previous section the metric in the near horizon limit in 10 dimensions is given by:
\be
ds^2_{10} = (-\frac{r^2}{R^2}dt^2 + \frac{R^2}{r^2}dr^2) + R^2d\Omega_2^2 + 2h_{m\bar n}dz^mdz^{\bar n}  
\ee
Here 
\bea
R^2 &=& (a_0c)^{1/2} \nonumber \\
h_{m\bar n} &=& -i(\frac{a_0}{c})^{1/2}(\omega_2)_{m\bar n} 
\eea
Thus the near horizon space time is AdS$_2\times $S$^2\times M'$.  The radius of curvature of both the AdS$_2$ and S$^2$ is $R$.  $M'$ is a six dimensional complex K{\" a}hler manifold whose metric is Ricci flat due to (\ref{id'}), i.e. $M'$ is a Calabi-Yau 3-fold.  The remaining supergravity fields are given as follows:
\bea
F' &=& \omega_2\wedge dvol(S^2) = R\Omega^{3} J_h\wedge dvol(S^2)\nonumber \\
G_2 &=& dC_1 = d(f^{-1}-1)= \Omega^{3}R^{-1}dt\wedge dr \\
e^{-\phi'} &=& \Omega^3 \nonumber 
\eea
where we have introduced the following notation: $\Omega = c^{1/12}a_0^{1/4}$ and $J_h$ is the K{\" a}hler form associated with the Calabi-Yau metric $h$.  This solution was also found in \cite{ssty} for $\Omega=1$.  

Let us go back to our original 11-dimensional space-time with an M5-brane wrapping a 4-cycle and an M-wave along the M5-brane transverse to the Calabi-Yau.  Using our results we find that the metric is:
\bea
ds^2_{11} &=& \Omega^{2}(-\frac{r^2}{R^2}dt^2 + \frac{R^2}{r^2}dr^2) + \Omega^{-4}(dy + (\frac{r}{a_0} -1)dt)^2
\nonumber \\
&+& \Omega^2R^2d\Omega_2^2 + 2k_{m\bar n}dz^mdz^{\bar n}
\eea
So the metric in d=11 is a product of S$^2\times$CY$_3$ with a U(1) bundle over AdS$_2$.  It is easy to check that the U(1) bundle is such that the 11-dimensional space-time is in fact AdS$_3\times$S$^2\times$CY$_3$.  The radius of curvature of AdS$_3$ is given by $\sqrt2\Omega R$, while the radius of curvature of S$^2$ is $\Omega R$.  The Calabi-Yau metric $k$ differs from $h$ by a factor of the dilaton:
\be
k_{m\bar n} = \Omega^2h_{m\bar n}
\ee

At this stage some comments are in order.  
\begin{itemize}
\item We have argued that the d=10 black hole space-time interpolates between R$^{(3,1)}\times$CY$_3$ and AdS$_2\times$S$^2\times$CY$_3$ (and in d=11 between R$^{(3,1)}\times$S$^1\times$CY$_3$ and AdS$_3\times$S$^2\times$CY$_3$).  These asymptotic and near-horizon geometries are products of simple factors.  The full metric which interpolates between these two regimes is not a product geometry.  The full metric is {\it not} even conformal to a product geometry.    The complexities of the geometry are hidden in the non-linear differential equations (\ref{eom}) and (\ref{f}) which simplify considerably in the two limiting regimes of the black hole space time.

\item The Calabi-Yau manifolds appearing in the the asymptotic and near-horizon regimes are not the same.   Consistent with the analysis of \cite{vafa} in the transition from a localized brane to flux through a cycle, the original manifold is replaced by a new one in which the minimum size of the cycle the flux threads is proportional to the charge of the original D-brane configuration.  In our case the localized charges of the D4-branes is replaced by a flux through a cycle $\Sigma_2$ which computes the number of D4-branes through the integral
\be
\int_{\Sigma_2\times S^2} F_4 = R^{-1}\Omega^3\int_{\Sigma_2}J_h \leq R^{-1}\Omega^3V_{\Sigma_2}
\ee 
The last inequality is a consequence of the fact that $J_h$ is a calibrating form for 2-cycles on the Calabi-Yau.  Similarly the number of D0-branes in the original configuration is computed in the near-horizon space-time by the integral
\be
\int_{CY_3\times S^2} *G_2 = \Omega^3R^{-1}V_{CY}
\ee
\item The attractor mechanism in the 4-dimensional supergravity theory tells us that the near-horizon solution is determined by the U(1) charges of the black hole - not by the asymptotic values of the scalar fields.  In our case this is explicit since the scalar fields are the K{\" a}hler moduli and specify the sizes of 2-cycles which are related directly to the flux and hence the charges of the black hole. 
 
\end{itemize}

\section{Black holes in four dimensions}
In this section we show how to find exact d=4 solutions using our formalism.  Here, we can go beyond the near horizon limit and find the full black hole geometry.  This section is based on ideas contained in \cite{kastor}, where the connection between the d=11 and d=5 descriptions of wrapped M5-branes was first explored.  Going from the N=2 solution in d=10 to d=4 is straightforward using a dimensional reduction ansatz.  When dimensionally reducing we assume that part of space-time can be treated as too small to be probed at energy scales we want to restrict to.  Thus we will truncate our d=10 metric given in (\ref{iia}) to four dimensions, assuming that the six dimensional part transverse to our 4 dimensions can be treated as a point at energy scales we are interested in\footnote{The solution we present in this section was also derived in \cite{behrndt} from a different point of view.  They treat their solution as a genuinely 10-dimensional supergravity solution.}  .  Hence:
\be
ds^2_{4} = -H^{-1/2}f^{-1/2}dt^2 + H^{1/2}f^{1/2}(dr^2 + r^2d\Omega^2) \label{bh4}
\ee
To be consistent with our treatment of the transverse space-time as a point we impose that:
\bea
\partial_m f &=& 0 \nonumber \\
\partial_m H &=& 0.
\eea
So (\ref{bh4}) is a genuine 4 dimensional metric.

The above ansatz simplifies our differential equations (\ref{eom}) and (\ref{f}) to
\bea
\frac{1}{r^2}\partial_r(r^2\partial_r g_{m\bar n}' )&=& 0 \nonumber \\
\frac{1}{r^2}\partial_r(r^2\partial_r f') &=& 0 \label{d4eom}
\eea
Notice that despite ignoring the transverse space we still have to solve for this metric to get a consistent truncation of the supergravity solution.  As before we take $J = ig_{m\bar{n}}dz^m\wedge dz^{\bar n}$.  We take the usual dimensional reduction ansatz consistent with the fact that $g$ is K{\" a}hler:
\be
J = \Phi^I(r) \alpha_I
\ee
where $\alpha_I$ are a basis of closed (1,1) forms.  The differential equations (\ref{d4eom}) become Laplace's equation in flat 3-dimensional space for the functions $\Phi^I$ and $f$.  These can be solved to yield:
\bea
\Phi^I &=& a^I + \frac{b^I}{r} \nonumber \\
f&=& 1 + \frac{a_0}{r}. \label{phi}
\eea
Where $a^I, a_0$, and $b^I$ are constants.  As $r\rightarrow\infty$ we move away from the black hole and therefore $J\rightarrow J_{CY} = a^I\alpha_I$.  Thus the metric on the Calab-Yau determines the constants $a^I$.  

We can find $H$ using equation (\ref{id}) to find
\be
H = \frac{1}{6V_{CY}}c_{IJK}\Phi^I\Phi^J\Phi^K
\ee
where the $c_{IJK}$ are the triple intersection numbers
\be
c_{IJK} = \int \alpha_I\wedge\alpha_J\wedge\alpha_K
\ee
and $V_{CY}$ is the volume of the Calabi-Yau
\be
V_{CY} = \frac{1}{6}c_{IJK}a^Ia^Ja^K.
\ee

The intrinsically four dimensional black hole is then given by the following which is, by construction, a solution to the N=2 supergravity theory in d=4.
\bea
ds^2 &=& -A^{-2} dt^2 + A^{2}(dr^2 + r^2 d\Omega_2^2) \nonumber \\\
e^{-\phi} &=& e^{-\phi_0}H^{-1/4}f^{3/4}
\eea
where 
\be
A^2 = H^{1/2}f^{1/2}.
\ee
The scalar partners of the (non-graviphoton) U(1) gauge fields in the four dimensional N=2 supergravity theory are given by \cite{kastor} $\Phi^I$ in (\ref{phi}) .  To complete our description of our black hole solution in 4 dimensions we give the appropriate 4-d U(1) field strengths as well.  From (\ref{F}) we see that using our ansatz for the metric we can write down an expression for the 4-form $F$ in d=10:
\bea
F &=& -r^2\partial_r\Phi^Idvol(S^2)\wedge \alpha_I \nonumber \\
&=& b^I dvol(S^2)\wedge \alpha_I. 
\eea
Defining $F_4 = F^I\wedge\alpha_I$ we can read off the U(1) field strengths in 4-dimensions associated with the reduction of the 3-form gauge potential on the Calabi-Yau:
\be
F^I = b^I dvol(S^2)
\ee
In addition to these gauge fields there is another U(1) gauge field coming from the 10-d R-R 1-form: $C_1$  The field strength associated with this U(1) gauge field is:
\be
G_2 = dC_1 = \frac{1}{a_0}dt\wedge dr
\ee
This completes our description of N=2 magnetically charged black holes descended from D4 and D0 branes.  This solution was also derived in \cite{behrndt} in a different way and treated as a 10-dimensional soluton.

We now briefly describe the near horizon geometry of these d=4 black holes.  It is easy to see that in the $r\rightarrow 0$ limit $\Phi^I \rightarrow b^I/r, f\rightarrow a_0/r$The geometry is AdS$_2\times$S$^2$ with radii of curvature of the two spaces given by 
\be
R^2= \sqrt{\frac{a_0c_{IJK}b^Ib^Jb^K}{6V_{CY}}}
\ee
The area of the horizon is $A= 4\pi R^2$.  The above formula was first derived in \cite{behrndt} and further explored in \cite{ms} and \cite{msw}.

\section{Generalizing the near horizon geometry to D0-D2-D4-D6 systems}
In this section we outline an approach to black hole near-horizon regimes that goes beyond the case of the wrapped D4/D0 system studied so far.   Our approach is as follows: we propose a general ansatz for all the supergravity fields based on certain expectations concerning properties of near horizon solutions.  We then impose the supergravity equations of motion to relate the free parameters of our ansatz.  We do not try to preserve any supersymmetry although it may be preserved in some cases. 

The ansatz will incorporate our prejudice that the near horizon geometry of black holes should be a product containing an AdS$_2\times$S$^2$ factor in d=10.  This is based on the existence of the same factor in near horizon geometries of black holes in d=4.  In addition we will assume that all supergravity fields are invariant under the isometries of AdS$_2\times$S$^2$.  

Our starting point is an ansatz for the near-horizon geometry in d=11 based on the expectations described above:
\be
ds^2 = R_1^2(-r^2dt^2 + \frac{dr^2}{r^2}) + R_2^2(d\theta^2 + sin^2\theta d\phi^2) + R_3^2(dy + A_idx^i)^2 + 2g_{m\bar n} dz^m dz^{\bar n}.
\ee
In the metric above, $x^i$ denote the coordinates on AdS$_2\times$S$^2$ and $R_1, R_2, R_3$ are constants.  The geometry in d=11 is thus a product of a complex manifold $M$ with metric $g_{m\bar n}$ and a U(1) bundle over AdS$_2\times$S$^2$.  The motivation for this ansatz is that it descends to a product of AdS$_2\times$S$^2$ and a complex manifold in d=10 after compactifying along $y$.  The reason we take the manifold $M$ to be complex is that all of the branes we consider here wrap holomorphic cycles in the Calabi-Yau, so we expect the complex structure of the Calabi-Yau to be preserved.  This ansatz hopefully encompasses a large class of, if not all, N=2 black holes.  

Next we present an ansatz for $F_4$ and $G_2 = dC_1 = d(A_idx^i)$.  We require that the entire supergravity solution respect the isometries of AdS$_2\times$S$^2$.  This leads us to the following ansatz:
\bea
F_4 &=& c_1 J\wedge\eta + c_2 J\wedge J + c_3dt\wedge dr\wedge\eta + c_4 dt\wedge dr \wedge J \nonumber \\
G_2 &=& a dt\wedge dr + b \eta \label{forms}
\eea
Here $\eta = \sin\theta d\phi\wedge d\theta = dvol(S^2)$ is the volume form of a unit sphere, and $J$ is the K{\" a}hler form associated with the metric $g$ on $M$.  The parameters $c_1, c_2, c_3, c_4, a, b$ are all constants.  There is a Killing symmetry generated by $\partial_y$, which means that we can make gauge transformations on $C_1= A_idx^i$ without affecting any physics.  We use this symmetry to pick the gauge:
\be
A_idx^i = ardt + b\cos\theta d\phi.
\ee
With this we have parameterized all of our fields.  We now turn to the equations of motion to find restrictions on the parameters of our ansatz.

We start with $F_4$.  We need to impose the following conditions:
\bea
dF_4 &=&0 \nonumber \\
d*F_4 &+& \frac{1}{2} F_4\wedge F_4 =0.
\eea
The first of these is the "Bianchi identity" and the second the equation of motion for $F_4$.  Depending on the case under study the role of these two equations can be interchanged.  In the absence of localized sources both equations appear on an equal footing as it does for us.  
Our ansatz (\ref{forms}) satisfies these equations only if:
\be
dJ = 0, \label{Kahler}
\ee
i.e. only if $g$ is a K{\" a}hler metric on $M$.  The requirement (\ref{Kahler}) also implies that 
\be
d*F=0
\ee
which in turn implies that $F_4\wedge F_4 =0$ or
\bea
c_1c_4+c_2c_3 &=& 0 \nonumber \\
c_1c_2 &=& 0 \\
c_2c_4 &=& 0\nonumber \label{cs}
\eea
In addition we need to impose:
\be
R_{ab} = \frac{1}{12}F_{acde}F_b^{\;\;cde} -\frac{1}{144}G_{ab}F_{cdef}F^{cdef}. \label{R}
\ee
The equations relate the parameters of our ansatz.  They are collected in the appendix.  We leave the full analysis of these relations for the future but we explain some general properties of solutions and look at a special solution below.  

\begin{itemize}
\item $M$ is a K{\"a}hler manifold
\item There are three branches solving (\ref{cs}).  Case (i) $c_2=c_1=0$, Case (ii) $c_2=c_4 =0$, Case (iii) $c_1=c_3=c_4=0$.   
\item The Ricci tensor for $M$ is
\be
R_{m\bar n} = \frac{1}{12}(\frac{9}{2}c_2^2 + 2c_3^2R_1^{-4}R_2^{-4})g_{m\bar n}
\ee
$M$ is therefore an Einstein manifold with a cosmological constant which is either 0 or positive.

\end{itemize}
It is straightforward to solve the equations in the appendix for the three cases described above.  We collect these results here and look at a special case in slightly more detail.  The appendix contains three equations which assume that $R_1, R_2, R_3$ are non-zero.  We will solve for three quantities in terms of the remaining undetermined quantities. 

{\bf Case (i)} ($c_2=c_1=0$) 
\bea
b^2&=&R_1^{-2}R_3^{-2}R_2^4+\frac{1}{2}R_1^{-4}R_2^{4}a^2 \nonumber \\
c_3^2&=& 2R_1^2R_2^2(R_2^2-R_1^2) \\
c_4^2&=&\frac{1}{3}R_1^2+\frac{2}{3}R_1^4R_2^{-2}- \frac{1}{2}R_3^2a^2\nonumber
\eea

{\bf Case(ii)}($c_2=c_4=0$)
\bea
a^2&=&R_2^{-2}R_3^{-2}R_1^4 +\frac{1}{2}R_2^{-4}R_1^{4}b^2 \nonumber \\
c_1^2&=&\frac{2}{3}R_1^{-2}R_2^4+\frac{1}{3}R_2^2-\frac{1}{2}R_3^2b^2 \\
c_3^2 &=& 2R_1^2R_2^2(R_2^2-R_1^2)\nonumber 
\eea

{\bf Case (iii)}($c_1=c_3=c_4=0$)
\bea
a^{2}&=&\frac{2}{3}R_3^{-2}R_1^4(2R_1^{-2}+R_2^{-2}) \nonumber \\
b^2&=& \frac{2}{3}R_3^{-2}R_2^4(R_1^{-2}+2R_2^{-2})\\
c_2^2&=& \frac{8}{9}(R_1^{-2}-R_2^{-2})\nonumber
\eea

We now present a special case which in type IIA corresponds to D2-branes wrapping 2-cycles and D6-branes wrapping the entire Calabi-Yau.  This case corresponds to only keeping $b$ and $c_4$ non-zero ($a=c_1=c_2=c_3=0$).  This falls into the category of Case (i) in the above classifcation.  $M$ is Ricci flat according to the analysis above.  It is straightforward to see from the equations presented in the appendix that $c_4^2=R_1^2 = R^2_2$, $b^2=R_3^{-2}R_1^2$.  The supergravity solution is then:
\bea
ds^2&=&R_1^2(-r^2dt^2 + \frac{dr^2}{r^2}) + R_1^2(d\theta^2 + \sin^2\theta d\phi^2) \nonumber \\
&+& R_3^2(dy + R_3^{-1}R_1\cos\theta d\phi)^2 + 2g_{m\bar n} dz^m dz^{\bar n}\nonumber \\
F_4 &=& R_1dt\wedge dr\wedge J \\
G_2 &=& R_3^{-1}R_1 \sin\theta d\phi\wedge d\theta\nonumber
\eea
This geometry is in fact AdS$_2\times$S$^3\times CY_3$ - the U(1) bundle over S$^2$ is such that it is S$^3$ with radius of curvature ${\sqrt 2}R_1$.  This should be contrasted with the D4/D0 brane case where the geometry was AdS$_3\times$S$^2\times CY_3$.  The U(1) bundle combines with the AdS$_2\times$S$^2$ base to give different products in the two cases. 

It is straightforward to write down completely general solutions for all three cases described above.  We leave a fuller treatment of these solutions including their brane interpretation for the future.  

\section{Conclusions and summary}
In this paper we gave an eleven and ten dimensional description of N=2 black holes which, from a ten dimensional perspective, arise from wrapped D4 and D0-branes.  We presented d=11 (and type IIA) supergravity equations which need to be solved to obtain the complete black hole solutions.  These equation are non-linear and are unlikely to be exactly solvable for arbitrary Calabi-Yau compactifications.  

These non-linear equations simplify considerably in the near-horizon limit.  We employ the ideas of large N geometric transitions advocated in \cite{vafa} to define a notion of near-horizon.  In this limit we find the general supergravity solution.  The geometry in d=10 is simply AdS$_2\times$S$^2\times CY_3$.  The lift of this near-horizon geometry to d=11 is AdS$_3\times$S$^2\times CY_3$.  

In section 5 we presented black hole solutions in d=4 starting with our 11/10 dimensional description of wrapped branes.  These solutions give the full black hole geometry (i.e. not just the near-horizon one).  

Finally, in section 6 we attempt to go beyond our supersymmetric wrapped D4/D0-brane system to include D2 and D6-branes as well.  We display an ansatz for all supergravity fields in the near-horizon regime consistent with the isometries of AdS$_2\times$S$^2$ and carrying all possible charges.  Our ansatz contains free parameters which are related to each other through the supergravity equations of motion.  These relations are collected in the Appendix.  We display a particular solution corresponding to wrapped D2 and D6 branes.  The near-horizon geometry in this case is a product AdS$_2\times$S$^3\times CY_3$.

There are many interesting directions to explore, especially in the context of recent exciting developments in d=4 N=2 black hole physics.  It would be interesting to see if our ansatz in section 6 is related to non-supersymmetric attractor solutions recently uncovered in \cite{non-susy}.  New developments in the area of black hole entropy for black holes arising in N=2 supergravity is another area where higher dimensional perspectives may provide insights.  
\no
\\

{\Large \bf Acknowledgements}\\
I am grateful to David Kastor for his very helpful comments on an earlier version of this paper as well as this one.  I thank Tasneem Zehra Husain for discussions.  I would also like to acknowledge funding from the Swedish VR. 
\\
\\
\no
{\Large\bf Appendix}\\ \\
In this section we present the equations of motion relevant for section 6.  
\bea
ds^2 &=& R_1^2(-r^2dt^2 + \frac{dr^2}{r^2}) + R_2^2(d\theta^2 + sin^2\theta d\phi^2) + R_3^2(dy + A_idx^i)^2 + 2g_{m\bar n} dz^m dz^{\bar n} \\
F_4 &=& c_1 J\wedge\eta + c_2 J\wedge J + c_3dt\wedge dr\wedge\eta + c_4 dt\wedge dr \wedge J \nonumber \\
G_2 &=& a dt\wedge dr + b \eta\nonumber
\eea
with $\eta = \sin\theta d\phi\wedge d\theta = dvol(S^2)$.  We now write down the conditions on the parameters $a,b, c_1, c_2, c_3, c_4, R_1, R_2, R_3$ arising from the equations of motion:
\be
R_{ij} = \frac{1}{12}F_{iklm}F_j^{\;\;klm} -\frac{1}{144}G_{ij}F_{klmn}F^{klmn}.
\ee
The relations between the parameters of our ansatz can all be stated very simply in three independent equations:
\bea
-1 +\frac{1}{2}R_3^2R_1^{-2}a^2 &=& -\frac{1}{2}c_1^2R_1^2R_2^{-4} -\frac{3}{8}c_2^2R_1^2-\frac{1}{3}c_3^2R_2^{-4}R_1^{-2}-c_4^2R_1^{-2}  \\
1-\frac{1}{2}R_3^2R_2^{-2}b^2 &=& c_1^2R_2^{-2}-\frac{3}{8}c_2^2R_2^2 -\frac{1}{3}c_3^2R_1^{-4}R_2^{-2}+ \frac{1}{2}c_4^2R_1^{-4}R_2^2 \\
R_3^2(R_2^{-4}b^2-R_1^{-4}a^2)&=&-c_1^2R_2^{-4}-\frac{3}{4}c_2^2+\frac{1}{3}c_3^2R_1^{-4}R_2^{-4}+c_4^2R_1^{-4}
\eea
In writing these relations I assumed that the parameters $R_i$ are {\it all} non-zero.  The other parameters $a,b,c_i$ are allowed to take on arbitrary values consistent with this assumption. 

The above equations can be analyzed in the context of our division of the parameters $c_i$ into three cases described in section 6.

}
\end{document}